\documentclass[journal=jctcce,manuscript=article]{achemso}
\setkeys{acs}{usetitle=true}

\usepackage[version=3]{mhchem} 

\usepackage{amsmath}
\usepackage{graphicx,color}
\usepackage{amsfonts}
\usepackage{epstopdf}
\usepackage{algorithm,algorithmic}
\usepackage{comment}

\usepackage[version=3]{mhchem} 

\newcommand{\Or}{\mathcal{O}}

\newcommand{\abs}[1]{\left\lvert#1\right\rvert}

\newcommand{\mc}[1]{\mathcal{#1}}

\newcommand{\ud}{\,\mathrm{d}}

\newcommand{\wt}[1]{\widetilde{#1}}

\newcommand{\hr}[1]{\hat{\vr}_{#1}}

\newcommand{\bvec}[1]{\mathbf{#1}}
\newcommand{\vr}{\bvec{r}}
\newcommand{\CY}[1]{{\color{cyan}~\textsf{[CY: #1]}}}

\newcommand{\REV}[1]{{\color{black}{#1}}}

\author{Wei Hu}
\email{whu@lbl.gov} \affiliation{Computational Research Division,
Lawrence Berkeley National Laboratory, Berkeley, California 94720,
United States}

\author{Lin Lin}
\email{linlin@math.berkeley.edu} \affiliation{Department of
Mathematics, University of California, Berkeley, California 94720,
United States} \alsoaffiliation{Computational Research Division,
Lawrence Berkeley National Laboratory, Berkeley, California 94720,
United States}

\author{Chao Yang}
\email{cyang@lbl.gov} \affiliation{Computational Research Division,
Lawrence Berkeley National Laboratory, Berkeley, California 94720,
United States}

\title[Density fitting for hybrid functional calculations]
{Interpolative Separable Density Fitting Decomposition for
Accelerating Hybrid Density Functional Calculations With
Applications to Defects in Silicon}

\begin{document}

\begin{abstract}

We present a new efficient way to perform hybrid density functional theory (DFT) based electronic structure calculation. The new method uses an interpolative separable density fitting (ISDF) procedure to construct a set of numerical auxiliary basis vectors and a compact approximation of the matrix consisting of products of occupied orbitals represented in a large basis set such as the planewave basis. Such an approximation allows us to reduce the number of Poisson solves from $\Or(N_{e}^2)$ to $\Or(N_{e})$ when we apply the exchange operator to occupied orbitals in an iterative method for solving the Kohn-Sham equations, where $N_{e}$ is the number of electrons in the system to be studied. We show that the ISDF procedure can be carried out in $\Or(N_{e}^3)$ operations, with a much smaller pre-constant compared to methods used in existing approaches. When combined with the recently developed adaptively compressed exchange (ACE) operator formalism, which reduces the number of times the exchange operator needs to be updated, the resulting ACE-ISDF method significantly reduces the computational cost \REV{associated with the exchange operator} by nearly two orders of magnitude compared to existing approaches for a large silicon system with $1000$ atoms. We demonstrate that the ACE-ISDF method can produce accurate energies and forces for insulating and metallic systems, and that it is possible to obtain converged hybrid functional calculation results for a 1000-atom bulk silicon within 10 minutes on 2000 computational cores. We also show that ACE-ISDF can scale to 8192 computational cores for a 4096-atom bulk silicon system. We use the ACE-ISDF method to geometrically optimize a 1000-atom silicon system with a vacancy defect using the HSE06 functional and computes its electronic structure. We find that that the computed energy gap from the HSE06 functional is much closer to the experimental value compared to that produced by semilocal functionals in the DFT calculations.


\end{abstract}

\section{Introduction} \label{sec:Introduction}

Kohn-Sham density functional theory
(KSDFT)\cite{PR_136_B864_1964_DFT, PR_140_A1133_1965_DFT} is the
most widely used electronic structure theory in condensed matter
physics and quantum chemistry.  The fidelity of the results produced
by a KSDFT calculation often depends on the choice
of the exchange and correlation functional.\cite{AIP_577_1_2001}
Hybrid exchange-correlation functionals, such as
B3LYP,\cite{JCP_98_1372_1993} PBE0\cite{JCP_105_9982_1996} and
HSE\cite{JCP_118_8207_2003,JCP_124_219906_2006_HSE06} are known to be more reliable in producing high fidelity results for a wide range of systems over
calculations that make use of local and semi-local exchange-correlation
functionals, such as the local density approximation
(LDA),\cite{PRL_45_566_1980, PRB_23_5048_1981, PRB_54_1703_1996_LDA}
the generalized gradient approximation
(GGA),\cite{PRA_38_3098_1988, PRB_37_785_1988, PRL_77_3865_1996_PBE}
and meta-GGA functionals.\cite{PRL_91_146401_2003_TPSS,
PRL_115_036402_2015_SCAN, PNAS_112_685_2015} However, hybrid
functionals include a fraction of the Fock exchange operator. Applying
$V_{X}[\{\psi_{i}\}](\vr,\vr') = -\sum_{i=1}^{n} \frac{\psi_{i}(\vr)
\psi_{i}(\vr')}{|\vr-\vr'|}$. Applying such an operator to a set of
$n$ orbitals $\Psi=[\psi_{1}(\vr),\ldots,\psi_{n}(\vr)]$, which
is often used in an iterative method for solving the Kohn-Sham
equations, requires solving
$\Or(N_{e}^2)$ Poisson-like equations, with effective charges taking
the form of $\psi_{i}(\vr)\psi_{j}(\vr) (1\le i,j\le n)$. Here $n\sim
\Or(N_e)$ and $N_{e}$ is the number of electrons. This is
costly, especially for calculations performed in a
large basis set such as planewaves and finite elements.
In these calculations, an iterative diagonalization procedure
is used solve the KS equations and the multiplication
of $V_{X}$ with $\Psi$, which has the complexity of $\Or(N_{e}^3)$
with a large pre-constant, needs to be performed in each iteration.
These multiplications alone often constitutes more than $95\%$ of
the overall computational time in a conventional approach.

There are two main routes to reducing the computational cost of
hybrid functional calculations. The first route is to reduce the
cost of multiplying $V_{X}$ with $\Psi$. This can be done through
efficient parallelization over a large number of
processors,\cite{DucheminGygi2010,BylaskaTsemekhmanBadenEtAl2011,CPC_214_52_2017_ACE}
or the use of linear scaling methods (i.e. $\Or(N_{e})$
methods).\REV{\cite{RMP_71_1085_1999_ON, RPP_75_036503_2010_ON,
JCTC_6_2348_2010, JCP_141_084502_2014, JCTC_11_4655_2015,
JCTC_11_1463_2015}} For large systems with a substantial band gap,
linear scaling methods use the nearsightedness
property\cite{PRL_76_3168_1996} to construct a sparse approximation
to the exchange operator, thereby reducing the cost of computing
$V_{X}\Psi$.\cite{JCTC_12_3514_2016} The second route is to reduce
the frequency of computing $V_{X}\Psi$,
and this route is much less explored until
recently\cite{JCP_143_024113_2015, BoffiJainNatan2016,
JCTC_12_2242_2016_ACE, JCTC_13_1188_2017_ACE}. The adaptively
compressed exchange (ACE) operator
formalism\cite{JCTC_12_2242_2016_ACE,JCTC_13_1188_2017_ACE} replaces
the dense and full rank exchange operator by a low rank operator
that is constructed on the fly. The low rank operator is only
updated once every few iterations. The ACE operator fully agrees
with $V_{X}$ in the subspace spanned by orbitals in $\Psi$. The
reduced rank of the ACE operator lowers the cost of $V_{X}\Psi$
calculation without losing accuracy. The ACE formulation is
applicable to insulators, semiconductors and metals. Recently we
have proved that such a low rank compression of the exchange
operator is uniquely determined through the ACE formulation, and
that the updating scheme for the ACE operator converges in the
self-consistent field (SCF) iteration both locally and globally for
the linearized Hartree-Fock-like equations.\cite{LinLindseyACE} The
ACE formulation can enable hybrid functional calculations in a
planewave basis set for more than a thousand
atoms,\cite{JCTC_13_1188_2017_ACE} and has recently been
integrated\cite{CPC_214_52_2017_ACE} into software packages
such as Quantum ESPRESSO.\cite{JPCM_21_395502_2009_QE}

In this paper, we develop a new method that combines the strength of
both approaches mentioned above to accelerate large scale hybrid
functional calculations. By using an interpolative separable density
fitting (ISDF) method first proposed by Lu and Ying in
Ref.~\cite{JCP_302_329_2015}, we construct a numerical auxiliary
basis for $\{\psi_{i}(\vr)\psi_{j}(\vr)\} (1\le i,j\le n)$ that
contains only $cN_e$ basis vectors for a small constant $c$. As a
result, applying $V_{X}$ to a set of $\Or(N_{e})$ orbitals only
requires solving $\Or(N_{e})$ instead of $\Or(N_{e}^2)$ Poisson-like
equations. Compared to the widely used density fitting
techniques\cite{ManzerHornMardirossianEtAl2015,
NJP_14_053020_2012,Weigend2002} in quantum chemistry, the main
feature of the ISDF decomposition is that the fitting coefficient
tensor, which is usually written as a three way tensor, can be
analytically separated into the product of two matrices. This is the
key to achieving $\Or(N_e^3)$ scaling, and avoiding the
$\Or(N_{e}^4)$ computational complexity that appears in many other
density fitting schemes. The ISDF decomposition is closely related
to the recently developed tensor hypercontraction (THC)
approach.\cite{ParrishHohensteinMartinezEtAl2012,
ParrishHohensteinMartinezEtAl2013}

\REV{The ISDF decomposition replaces $\{\psi_{i}(\vr)\psi_{j}(\vr)\}
(1\le i,j\le n)$ with the product of two matrices. One of the
matrices, which can be viewed as a matrix of fitting coefficients,
simply consists of $\{\psi_{i}\psi_{j}\}$ evaluated at a set of
carefully chosen \textit{interpolation points} $\hat{r}_{\mu}$, for
$\mu = 1,...,N_{\mu}$ and $N_{\mu} = cN_e$. The other matrix
contains numerical auxiliary basis vectors that we will also refer
to as the \textit{interpolating vectors}. In this paper, these two
matrices are determined separately. The matrix containing the
fitting coefficients, which depends solely on the choice of
interpolation points, is determined first. The matrix containing the
interpolating vectors is subsequently obtained through a least
squares fitting procedure. This approach is different from the
decomposition proposed in Ref.~\cite{JCP_302_329_2015}, where the
numerical auxiliary basis and the fitting coefficients are
determined simultaneously through a randomized QR factorization with
column pivoting (QRCP) applied to $\{\psi_{i}(\vr)\psi_{j}(\vr)\}$
directly. We find that using randomized QRCP at each SCF iteration
is costly, and does not speed up hybrid functional calculations. By
separating the treatment of the interpolation points from the
construction of the matrix containing the interpolating vectors in
ISDF, we can use the relatively expensive randomized QRCP procedure
to find the interpolation points in advance, and only recompute the
interpolation vectors whenever $\{\psi_{i}(\vr)\psi_{j}(\vr)\}$ has
been updated using an efficient least squares procedure that
exploits the separable nature of the matrix to be approximated. As a
result, we can significantly accelerate hybrid functional
calculations using the ISDF decomposition in all but the first SCF
iteration.  }

\REV{The ISDF decomposition can be used in the construction of
ACE operator to reduce the number of Poisson solves required
in the construction. In fact, the decomposition itself yields
a low-rank approximation of the Fock exchange operator. However,
symmetry is not strictly preserved in the ISDF decomposition.
The lack of symmetry can introduce numerical stability issues
in the convergence of the SCF iteration.
We will demonstrate how to combine the ISDF decomposition with the ACE
formulation in a numerically stable manner by maintaining the symmetry
of the compressed exchange operator.} The resulting ACE-ISDF method does not
rely on the nearsightedness property, and is applicable to
insulators, semiconductors and metals. The computational complexity
of our new approach is still $\Or(N_{e}^{3})$, but the pre-constant
is significantly reduced. The ACE-ISDF method can be efficiently
parallelized on high performance supercomputers. Using this
technique, we can perform hybrid functional calculations for a bulk
silicon system with 1000 atoms in less than 10 wall clock minutes on
2000 computational cores. \REV{We find that the cost associated with the
exchange operator is reduced by nearly two orders of magnitude compared
to conventional approaches.} Furthermore, this method can
also scale to 8192 computational cores for a 4096-atom bulk silicon
system.


As an example, we use the ACE-ISDF method to optimize the geometry
and compute the electronic structure of a bulk 1000-atom silicon
system that contains a single vacancy, at the level of the HSE06
hybrid functional calculations~\cite{JCP_124_219906_2006_HSE06}.
Our calculation reveals three defect states within the intrinsic
energy gap of the silicon. The computed energy gap is much closer to
the experimental value compared to GGA functional calculations.


The rest of the paper is organized as follows. In
section~\ref{sec:prelim} we introduce the density fitting approximation
in the context of hybrid functional calculations, and the interpolative
separable density fitting approximation.
In section~\ref{sec:isdfhybrid} we develop a new method to
efficiently compute the interpolative separable density fitting
approximation, and to combine with the
adaptively compressed exchange formulation. We describe an efficient
parallelization strategy in section~\ref{sec:Parallel}. The numerical
results are given in section~\ref{sec:Result}, followed by conclusion
and discussion in section~\ref{sec:conclusion}.




\section{Preliminaries}\label{sec:prelim}

\subsection{Density fitting approximations in hybrid functional calculations}\label{sec:hybrid}

For simplicity, we consider isolated, gapped systems and omit
spin degeneracy.
In hybrid functional calculations, the exchange operator is an integral
operator defined in terms of the occupied orbitals
$\{\varphi_{i}\}_{i=1}^{N_{e}}$ with kernel
\begin{equation}
  V_{X}[\{\varphi_{i}\}](\vr,\vr') =
  -\sum_{i=1}^{N_{e}} \varphi_{i}(\vr) \varphi_{i}(\vr') K(\vr,\vr')
  \equiv
  -P^{\varphi}(\vr,\vr')K(\vr,\vr'),
  \label{eqn:VXkernel}
\end{equation}
where $K$ is either the Coulomb potential
$K(\vr,\vr')=\frac{1}{\abs{\vr-\vr'}}$ in hybrid functionals such as
B3LYP\cite{JCP_98_1372_1993} and PBE0,\cite{JCP_105_9982_1996} or
the screened Coulomb potential
$K(\vr,\vr')=\frac{\mathrm{erfc}(\mu|\vr-\vr'|)}{|\vr-\vr'|}$ in
functionals such as HSE.\cite{JCP_118_8207_2003} $P^{\varphi}$
is the density matrix.

When a large basis set such as the planewave basis set is used to
discretize the KS equations, it is generally more efficient to
perform $V_{X}[\{\varphi_{i}\}]\psi_{j}$, $j=1,2,...,n$ on the fly in
an iterative diagonalization procedure without explicitly
constructing or storing $V_{X}[\{\varphi_{i}\}]$. In many cases,
$n=N_{e}$, but $\{\varphi_{i}\}$ and $\{\psi_{i}\}$ may be different
sets of orbitals when used in a self-consistent field (SCF)
iteration.  It is also possible to have $n>N_{e}$ when some
unoccupied orbitals are also to be computed. So we deliberately use
different notation  and to distinguish the occupied orbitals
$\{\varphi_{i}\}$ from generic orbitals $\{\psi_{i}\}$ that can be
either occupied or unoccupied. In order to reach self consistency
for the occupied orbitals $\{\varphi_{i}\}$, a common practice is to
separate the self-consistent field (SCF) iteration into two sets of
SCF iterations. In the \textit{inner SCF iteration}, the exchange operation
$V_{X}$ defined by the orbitals $\{\varphi_{i}\}$ is fixed and the
Hamiltonian operator only depends on the density $\rho(\vr)$. The
SCF iteration then proceeds as in KSDFT calculations with a
fixed exchange operator. In the \textit{outer SCF iteration}, the occupied
components of the output orbitals $\{\psi_{j}\}$ can be used as the
input orbitals to update the exchange operator. In each inner
iteration, the product of $V_{X}[\{\varphi_{i}\}]$ and $\psi_{j}$
need to be evaluated many times using the relation
\begin{equation}
  \left(V_{X}[\{\varphi_{i}\}]\psi_{j}\right)(\vr) =
  -\sum_{i=1}^{N_{e}} \varphi_{i}(\vr) \int K(\vr,\vr')
  \varphi_{i}(\vr')\psi_{j}(\vr') \ud \vr'.
  \label{eqn:applyVX}
\end{equation}
The integration in
Eq.~\eqref{eqn:applyVX} is often carried out by solving Poisson-like
equations, using e.g. a fast Fourier transform (FFT) method.
The number of equations to be solved is $n N_{e}\sim \Or(N_{e}^2)$. This
is typically the most time consuming component in hybrid functional
calculations.

We would like to reduce the number of equations to be solved by
exploiting the numerical rank deficiency in the set of right-hand
sides $\{\varphi_{i}(\vr) \psi_{j}(\vr)\}$ in these Poisson-like
equations, and representing them using a smaller set of linearly
independent basis. One possible way to achieve this is through the use
of a density fitting method (a.k.a. resolution of
identity).\cite{Weigend2002,NJP_14_053020_2012} In general, given
two sets of functions $\{\varphi_{i}(\vr)\}_{i=1}^{m}$,
$\{\psi_{j}(\vr)\}_{j=1}^{n}$, a density fitting procedure
constructs an auxiliary basis $\{\zeta_\mu\}$, $\mu =
1,2,...,N_{\mu}$, with $N_{\mu} \ll mn$, for the set of Hadamard
products (i.e. the element-wise product)
\begin{equation}
\{Z_{ij}(\vr) :=\varphi_{i}(\vr)\psi_{j}(\vr)\}_{1\le i\le m,1\le j\le n}
\label{eq:zij}
\end{equation}
so that
\begin{equation}
 \varphi_{i}(\vr)  \psi_{j}(\vr)\approx \sum_{\mu=1}^{N_{\mu}}
\zeta_{\mu}(\vr) C_{\mu}^{ij},
  \label{eqn:dfformat}
\end{equation}
where $C_{\mu}^{ij}$'s are fitting coefficients. This can be
implemented using rank revealing methods such as the singular value
decomposition (SVD),\cite{NumerischeMathematik_14_403_1970} and the
pivoted Cholesky
factorization.\cite{AppliedNumericalMathematics_62_428_2012}

In the context of hybrid functional calculations above, we have
$m=N_{e}$ and $n\sim \Or(N_{e})$. Hence, the density fitting
procedure compresses $mn\sim \Or(N_{e}^2)$ functions into a much
smaller set of auxiliary functions
$\{\zeta_{\mu}(\vr)\}_{\mu=1}^{N_{\mu}}$. Numerical results indicate
that it is often sufficient to choose $N_{\mu}=c N_{e}$, where $c$
is a small constant that we refer to as a rank parameter. This
parameter determines the computational accuracy of the
decomposition~\eqref{eqn:dfformat}.

In the standard density fitting procedure, the fitting coefficient tensor
$\{C_{\mu}^{ij}\}$ is treated as a three way tensor, and is often
obtained through a least squares fitting procedure. The storage cost
of $\{C_{\mu}^{ij}\}$ is $\Or(N_{e}^3)$ and the computational cost
of density fitting typically scales as $\Or(N_{e}^{4})$. When a
large basis set such as the planewave basis set is used, both the
storage and the computational cost can be prohibitively high. As a
result, density fitting is rarely used for this type of basis set
unless additional locality constraints are
enforced.\cite{NJP_14_053020_2012}

\subsection{Interpolative separable density fitting decomposition}

In order to reduce the complexity of the density fitting method, the
key is to find a more efficient treatment for the three way fitting
coefficient tensor. This has been achieved by the tensor
hypercontraction (THC)
method,\cite{ParrishHohensteinMartinezEtAl2012,ParrishHohensteinMartinezEtAl2013}
and the interpolative separable density fitting (ISDF)
method.\cite{JCP_302_329_2015} Both methods use the following
compression format
\begin{equation}
  \varphi_{i}(\vr)  \psi_{j}(\vr)\approx \sum_{\mu=1}^{N_{\mu}}
  \zeta_{\mu}(\vr) \varphi_{i}(\hr{\mu})\psi_{j}(\hr{\mu}).
  \label{eqn:isdfformat}
\end{equation}
Here $\{\hr{\mu}\}_{\mu=1}^{N_{\mu}}$ is a subset of real space grid
points $\{\vr_{i}\}_{i=1}^{N_{g}}$ on which the orbitals are
evaluated. We will refer to  $\{\hr{\mu}\}_{\mu=1}^{N_{\mu}}$ as the
interpolation points, and $\{\zeta_{\mu}(\vr)\}_{\mu=1}^{N_{\mu}}$
sampled on $\{\vr_{i}\}_{i=1}^{N_{g}}$  the interpolation vectors.
Since the term ``interpolative separable'' captures clearly the
relation between the format~\eqref{eqn:isdfformat} and standard
density fitting formats, with some abuse of terminology, we will
also refer the format~\eqref{eqn:isdfformat} as the ``ISDF format'' or
the ``ISDF decomposition''.

Comparing Eq.~\eqref{eqn:dfformat} with Eq.~\eqref{eqn:isdfformat},
we see that the ISDF decomposition is a special form of a density
fitting decomposition, where the fitting coefficient tensor
$C_{\mu}^{ij} =\varphi_{i}(\hr{\mu})\psi_{j}(\hr{\mu})$ is given
explicitly \textit{without additional computation}. Hence the
storage cost is reduced from $\Or(N_{e}^{3})$ to $\Or(N_{e}^2)$, and
the ISDF decomposition is potentially suitable for calculations with
a large basis set. The reason why such decomposition can be expected
can be understood from the perspective of interpolation. Indeed, if
$\{\hr{\mu}\}$ is a set of grid points in the real space, and let
$\zeta_{\mu}(\vr)$ be the Lagrange interpolation function on these
grid points satisfying
\begin{equation}
  \zeta_{\mu}(\hr{\mu'}) = \begin{cases}
    1, & \mu = \mu',\\
    0, & \mathrm{otherwise}
  \end{cases},
  \label{}
\end{equation}
then the ISDF decomposition would become sufficiently accurate as
one systematically refines the set $\{\hr{\mu}\}_{\mu=1}^{N_{\mu}}$.
In the worst case, all grid points are selected and $N_{\mu}=N_{g}$.
Since $N_{g}\sim \Or(N_{e})$, the asymptotic number of interpolation
vectors is still smaller than $\Or(N_{e}^2)$ even in such scenario.
Furthermore, if both $\{\varphi_{i}(\vr)\}$ and $\{\psi_{i}(\vr)\}$
are sets of sufficiently smooth functions, the number of points
$N_{\mu}$ can be expected to be much smaller than $N_{g}$.

The THC and ISDF methods differ in terms of the procedure for
finding the interpolation points and interpolation vectors, and
hence the computational complexity. For the THC method, the
interpolation points are determined through a quadrature rule, e.g.
uniform grid points, or grid points from a Gauss-Hermite quadrature
rule. However, the focus of THC is not to identify the interpolation
vectors, but to find an efficient approximation scheme for the four
way Coulomb integral tensor, which can be written in the current
context as
\begin{equation}
  V_{ijkl} = \int K(\vr,\vr')
  \varphi_{i}(\vr)\psi_{j}(\vr)\varphi_{k}(\vr') \psi_{l}(\vr')\ud \vr
  \ud \vr'.
  \label{eqn:fourtensor}
\end{equation}
Using the ISDF format, we have
\begin{equation}
  V_{ijkl} \approx
  \sum_{\mu,\nu=1}^{N_{\mu}}
  \varphi_{i}(\hr{\mu})\psi_{j}(\hr{\mu}) M_{\mu,\nu}
  \varphi_{k}(\hr{\nu})\psi_{l}(\hr{\nu}).
  \label{eqn:VijklISDF}
\end{equation}
Here the coefficient matrix is given by the interpolation vectors
\begin{equation}
  M_{\mu,\nu} = \int K(\vr,\vr') \zeta_{\mu}(\vr) \zeta_{\nu}(\vr') \ud
  \vr \ud \vr'.
  \label{eqn:THCcoef}
\end{equation}
On the other hand, if $V_{ijkl}$ is already known, it is possible to not
to explicitly evaluate the interpolation vectors, but directly obtain
$M_{\mu,\nu}$ using the following equation
\begin{equation}
  \begin{split}
  &\sum_{ijkl} \varphi_{i}(\hr{\mu}) \psi_{j}(\hr{\mu})V_{ijkl}
  \varphi_{k}(\hr{\nu}) \psi_{l}(\hr{\nu}) \\
  \approx & \sum_{\mu',\nu'=1}^{N_{\mu}} \left(\sum_{i}
  \varphi_{i}(\hr{\mu}) \varphi_{i}(\hr{\mu'})\right)
  \left(\sum_{j}
  \psi_{j}(\hr{\mu}) \psi_{j}(\hr{\mu'})\right)
  M_{\mu',\nu'}
  \left(\sum_{k}
  \varphi_{k}(\hr{\nu}) \varphi_{k}(\hr{\nu'})\right)
  \left(\sum_{l}
  \psi_{l}(\hr{\nu}) \psi_{l}(\hr{\nu'})\right),
  \end{split}
  \label{eqn:THCleastsquare}
\end{equation}
Eq.~\eqref{eqn:THCleastsquare} can be understood as a least squares
fitting to the equation~\eqref{eqn:VijklISDF}. The dominant
computational cost comes from the evaluation of the left-hand side,
which scales as $\Or(N_e^{5})$ if we do not assume any structure on
the four way tensor $V_{ijkl}$. Hence the strategy of THC is not
suitable for large basis set calculations.

The ISDF method~\cite{JCP_302_329_2015} uses a different
approach to construct the decomposition~\eqref{eqn:isdfformat}.
Instead of using a quadrature rule, it uses a randomized QR
factorization with column pivoting (QRCP)
procedure\cite{SIAM_13_727_1992_QRCP} to find the interpolation
points for the product pairs
$\{\varphi_{i}(\vr)\psi_{j}(\vr)\}$, which can be potentially much
more compact than a set of universal quadrature points. The
interpolation vectors can also be deduced from the QRCP
decomposition in the same calculation. The computational cost of the
ISDF decomposition is only $\Or(N_{e}^3)$, and hence is suitable for
large basis set calculations. It should be pointed out that the ISDF
decomposition yields a decomposition for the Coulomb integral
tensor once the coefficient matrix in Eq.~\eqref{eqn:THCcoef} is
computed, but the reverse statement is not true. Some
of the key steps of the randomized QRCP procedure will be discussed
in section~\ref{sec:isdfhybrid}.

\begin{figure}[htbp]
\begin{center}
\includegraphics[width=0.5\textwidth]{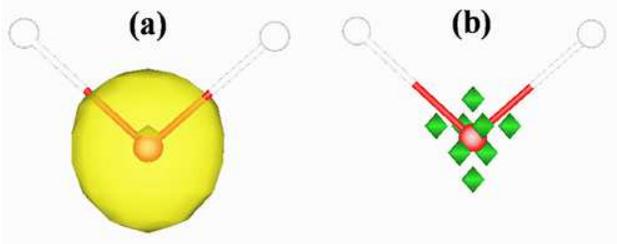}
\end{center}
\caption{(Color online) (a) The electron density (yellow
isosurfaces) and (b) the interpolation
points (green squares) $\{\hr{\mu}\}_{\mu=1}^{N_{\mu}}$ ($N_{\mu}$ =
8) selected from the real space grid points
$\{\vr_{i}\}_{i=1}^{N_{g}}$ ($N_{g}$ = 66$^3$) for a water molecule in
a 10 {\AA} $\times$ 10 {\AA} $\times$ 10 {\AA} box. The white and
red balls denote hydrogen and oxygen atoms, respectively.} \label{fig:H2O}
\end{figure}

As an example, we consider a water molecule (four occupied bands
$N_\text{band}$ = 4) (Figure~\ref{fig:H2O} (a)) in a 10 {\AA} $\times$ 10
{\AA} $\times$ 10 {\AA} box, with the kinetic energy cutoff
$E_\text{cut}$ = 60 Hartree. This corresponds to a real space grid of
size $66\times 66\times 66$ (i.e. $N_{g}$ = 66$^3$).
\REV{Figure~\ref{fig:H2O} (b) shows how $N_{\mu}=8$ interpolation points are
distributed in the real space.} As we can see, they are closer to the
oxygen atom than to hydrogen atoms. This is consistent with the distribution
of the electron density.
Figure~\ref{fig:H2O} (b) also indicates that instead of using a uniform
sampling grid, it may be more advantageous to select the interpolation
points adaptively from QRCP, especially for electron densities
with an inhomogeneous spatial distribution.

\section{ISDF decomposition for hybrid functional
calculations}\label{sec:isdfhybrid}

Although the ISDF decomposition significantly reduces the number of
Poisson-like equations to be solved, the complexity of the
randomized QRCP method used to find the interpolation points is still
$\Or(N_{e}^{3})$, which is comparable to the cost of computing
$V_{X}[\{\varphi_{i}\}]\psi_{j}$.
Hence at first glance the ISDF decomposition may not lead to
much efficiency gain in the context of hybrid functional
calculations.

Below we introduce a new method to compute the ISDF decomposition, which
separates the treatment of the interpolation vectors and the
interpolation points.  More specifically,  we use the randomized QRCP
decomposition only to find the interpolation points, and use a least
squares fitting procedure to efficiently find the interpolation vectors
without randomization.
Although it is more expensive to use the
randomized QRCP decomposition, this step only needs to be invoked
once in the entire hybrid functional calculation. The calculation of
the interpolation vectors still scales as $\Or(N_{e}^3)$, but the
pre-constant is significantly smaller.  We find that such choice
balances the efficiency and accuracy, especially when the
decomposition needs to be repeatedly used such as in a SCF iteration
procedure.  We also discuss how to combine the ISDF decomposition
with the adaptively compressed exchange operator (ACE) formulation
in a numerically stable way, so that the overall computational cost
can be significantly reduced.

\subsection{Finding the interpolation vectors}\label{subsec:interpvector}

We first discuss how to find the interpolation vectors assuming that
the interpolation points $\{\hr{\mu}\}$ are given.
%

Note that Eq.~\eqref{eqn:isdfformat} can be written as
\begin{equation}
  Z - \Theta C = 0,
  \label{eqn:isdflineq}
\end{equation}
where each column of $Z$ is defined by Eq.~\eqref{eq:zij} sampled on
real space grids $\{\vr_{i}\}_{i=1}^{N_g}$. $\Theta = [\zeta_1,
\zeta_2, ..., \zeta_{N_{\mu}}]$ contains the interpolating vectors,
and the column of $C$ indexed by $(i,j)$ is given by
\[ [ \varphi_{i}(\hr{1})\psi_{j}(\hr{1}), \cdots,
    \varphi_{i}(\hr{\mu})\psi_{j}(\hr{\mu}), \cdots,
    \varphi_{i}(\hr{N_{\mu}})\psi_{j}(\hr{N_{\mu}})]^T.
\]
Eq.~\eqref{eqn:isdflineq} is an over-determined linear systems with
respect to the interpolation vectors $\Theta$. One possible way to
solve the over-determined system is to impose the Galerkin condition
\begin{equation}
(Z - \Theta C) C^T = 0.
\label{eq:galerkin}
\end{equation}
It follows that the interpolating vectors can be obtained from
\begin{equation}
\Theta = ZC^T (CC^T)^{-1}. \label{eq:Theta}
\end{equation}
Note that the solution given by Eq.~\eqref{eq:Theta} is a least
squares approximation to the solution of Eq.~\eqref{eqn:isdfformat}.
This is similar to that in the THC method, but the important
difference is that the least squares fitting is applied to the $Z$
matrix, which is the key to reduce the complexity.

It may appear that the matrix-matrix multiplications $ZC^T$ and
$CC^T$ take $\Or(N_{e}^{4})$ operations because the size of $Z$ is
$N_g \times (N_{e}n)$ and the size of $C$ is $N_{\mu} \times
(N_{e}n)$.  However, both multiplications can be carried out with
fewer operations due to the separable structure of $Z$ and $C$. It
follows from the identity
\[ \sum_{i,j} \varphi_{i}\psi_{j} =
\left(\sum_{i} \varphi_{i} \right) \left( \sum_{j} \psi_{j} \right)
\]
that we may rewrite the $(k,\mu)$th element of $ZC^T$ (also denoted by
$P^{\varphi\psi}$) as
\begin{equation}
P^{\varphi\psi} \equiv
e_k^TZC^Te_{\mu} = P^{\varphi}(\vr_k,\hr{\mu}) P^{\psi}(\vr_k,\hr{\mu}),
\label{eq:newzct}
\end{equation}
where $P^{\varphi}(\vr,\hat{\vr}_{\mu})$ and
$P^{\psi}(\vr,\hat{\vr}_{\mu})$ can be
viewed as columns of (quasi) density matrices defined as
\begin{equation}
  P^{\varphi}(\vr,\vr') =
  \sum_{i=1}^{m}\varphi_i(\vr)\varphi_i(\vr'), \quad
  P^{\psi}(\vr,\vr') =
  \sum_{j=1}^{n}\psi_j(\vr)\psi_j(\vr').
  \label{eqn:quasidm}
\end{equation}
Similarly, we can rewrite the $(\nu,\mu)$th element of $CC^T$ as
\begin{equation}
e_{\nu}^TCC^T e_{\mu} =  P^{\varphi}(\hr{\nu},\hr{\mu}) P^{\psi}(\hr{\nu},\hr{\mu})
\label{eq:newcct}
\end{equation}

Because both $P^{\varphi}$ and $P^{\psi}$ matrices can be evaluated
with  $\Or(N_{e}^{3})$ floating point operations, and the
multiplications in Eq.~\eqref{eq:newzct} and Eq.~\eqref{eq:newcct}
are pointwise multiplications (Hadamard products) consuming only
$\Or(N_{e}^{2})$ floating point operations, the computational complexity for computing
the interpolation vectors is $\Or(N_{e}^{3})$.

The ISDF decomposition can be readily used to accelerate the
computation in Eq.~\eqref{eqn:applyVX}. Substituting
Eq.~\eqref{eqn:isdfformat} into Eq.~\eqref{eqn:applyVX}
 yields
\begin{equation}
\begin{split}
  \left(V_{X}[\{\varphi_{i}\}]\psi_{j}\right)(\vr)
  &= -\sum_{i=1}^{N_{e}} \left( \varphi_{i}(\vr) \int
  K(\vr,\vr') \varphi_{i}(\vr')\psi_{j}(\vr') \ud
  \vr' \right) \\
  &\approx -\sum_{i=1}^{N_{e}} \sum_{\mu=1}^{N_{\mu}} \varphi_{i}(\vr)
  \left(\int K(\vr,\vr') \zeta_{\mu}(\vr') \ud \vr'\right)
  \varphi_{i}(\hr{\mu}) \psi_{j}(\hr{\mu})\\
  &\equiv -\sum_{\mu=1}^{N_{\mu}} P^{\varphi}(\vr,\hr{\mu}) V^{\zeta}_{\mu}(\vr)
  \psi_{j}(\hr{\mu}),
\end{split}
\label{eqn:applyVXDF}
\end{equation}
where $P^{\varphi}(\vr,\hr{\mu})$ is given by the quasi density matrix
defined in Eq.~\eqref{eqn:quasidm}, and
\begin{equation}
  V^{\zeta}_{\mu}(\vr) \equiv \int K(\vr,\vr') \zeta_{\mu}(\vr') \ud \vr'
  \label{eqn:Vzeta}
\end{equation}
can be carried out as the solution to a Poisson-like equation. The
ISDF decomposition reduces the total number of Poisson-like
equations to be solved from $\Or(N_{e}^2)$ to $N_{\mu}\sim
\Or(N_{e})$.

\subsection{Finding the interpolation points}\label{subsec:interppoint}

The problem for finding a suitable set of interpolation points
$\{\hr{\mu}\}_{\mu=1}^{N_{\mu}}$ can be formulated as the following
linear algebra problem. Consider the discretized matrix $Z$ of size
as an $N_{g}\times (mn)$ matrix $Z$, and find $N_{\mu}$ rows of $Z$
so that the rest of the rows of $Z$ can be approximated by the
linear combination of the selected $N_{\mu}$ rows. This is called an
interpolative decomposition\cite{SIAM_13_727_1992_QRCP}, and a
standard method to achieve such a decomposition is the QR
factorization with column pivoting (QRCP)
procedure\cite{SIAM_13_727_1992_QRCP} as
\begin{equation}
  Z^{T} \Pi = QR,
  \label{eqn:QRCP}
\end{equation}
where $Z^T$ is the transpose of $Z$, $Q$ is an $mn \times N_g$
matrix that has orthonormal columns, $R$ is an upper triangular
matrix, and $\Pi$ is a permutation matrix chosen so that the
magnitude of the diagonal elements of $R$ form an non-increasing
sequence.  The magnitude of each diagonal element $R$ indicate how
important the corresponding column of the permuted $Z^T$ is, and
whether the corresponding grid point should be chosen as an
interpolation point. The QRCP factorization can be terminated when
the $(N_{\mu}+1)$-th diagonal element of $R$ becomes less than a
predetermined threshold. The leading $N_{\mu}$ columns of the
permuted $Z^T$ are considered to be linearly independent
numerically. The corresponding grid points are chosen as the
interpolation points. The indices for the chosen interpolation
points $\{\hr{\mu}\}$ can be obtained from indices of the nonzero
entries of the first $N_{\mu}$ columns of the permutation matrix
$\Pi$. However, the storage requirement for the matrix $Z$ is
$\Or(N_{e}^{3})$ and the computational cost associated with a
standard QRCP procedure is $\Or(N_{e}^{4})$, which is not so
appealing.

The key idea used in Ref.~\cite{JCP_302_329_2015} to lower the cost of
QRCP is to use a random matrix to subsample columns of the matrix
$Z$ to form a smaller matrix $\wt{Z}$ of size $N_{g}\times
\wt{N}_{\mu}$, where $\wt{N}_{\mu}$ is only slightly larger than
$N_{\mu}$. It can be shown that under some mild assumptions, the
reduction in the number of columns in a randomly subsampled $Z$ does
not have much impact the quality of the interpolation points
$\{\hr{\mu}\}$. However, we will not present the theoretical
analysis here, but merely describe the algorithmic ingredients. We
refer readers to Ref.~\cite{PNAS_104_20167_2007,
SIAMRev_53_217_2011} for more detailed analysis of randomized
sampling methods.

There are a number of ways to subsample columns of $Z$. Instead of using
the subsampled Fourier transform as in Ref.~\cite{JCP_302_329_2015},
here we choose two orthogonalized Gaussian matrices $G^{\varphi},G^{\psi}$ of size
$m\times p$ and $n\times p$, respectively, where $p\sim
\Or(\sqrt{N_{\mu}})$ is chosen to satisfy $\wt{N}_{\mu}=p^2$, and use
them to construct a set of subsampled products defined by
\begin{equation}
  \wt{Z}_{\alpha\beta}(\vr) = \left(\sum_{i=1}^{m} \varphi_{i}(\vr)
  G^{\varphi}_{i\alpha}\right)\left(\sum_{j=1}^{n} \psi_{j}(\vr)
  G^{\psi}_{j\beta}\right), \quad 1\le \alpha,\beta\le p.
  \label{eqn:compressZ}
\end{equation}
The corresponding discretized matrix $\wt{Z}$ is of size $N_{g}\times
\wt{N}_{\mu}$. Applying the QRCP procedure
to $\wt{Z}$ yields
\begin{equation}
  \wt{Z}^{T} \Pi = QR,
  \label{eqn:QRCPZt}
\end{equation}
where the interpolation points $\{\hr{\mu}\}_{\mu=1}^{N_{\mu}}$ are
given by the first $N_{\mu}$ columns of the permutation matrix
$\Pi$. Since the random matrices $G^{\varphi}$ and $G^{\psi}$ are
only applied to $\{\varphi_{i}\}$ and $\{\psi_{j}\}$ respectively,
the storage cost for $\wt{Z}$ is $\Or(N_{e}^2)$, and the
computational cost for generating $\wt{Z}$, which is dominated by
the cost of matrix-matrix multiplications is $\Or(N_{e}^{2.5})$. The
reduced matrix size allows the computational cost of the QRCP
procedure to be reduced to $\Or(N_{e}^{3})$. Since the QRCP
algorithm has been implemented in standard linear algebra software
packages such as LAPACK and ScaLAPACK\cite{ScaLAPACK_2011},  the
implementation and parallelization of ISDF is relatively
straightforward.

Our numerical results indicate that the cost of the randomized QRCP
method can be comparable to that of computing
$V_{X}[\{\varphi_{i}\}]\psi_{j}$. However, while the interpolation
vectors depends sensitively on the shape of the input orbitals
$\{\varphi_{i}\}$ and $\{\psi_{j}\}$ and need to be recomputed
whenever they are updated, the interpolation points are much less
sensitive to small changes of the orbitals. This is because the
significance of the interpolation points is only to indicate which
columns of $Z^{T}$ are important. In practice we find it sufficient
to determine the interpolation points at the beginning of the hybrid
functional calculations starting from a set of KS orbitals from a
converged GGA calculation, and to use such interpolation points
throughout the SCF iterations.


\subsection{Combining with the adaptively compressed exchange operator formulation}
\label{sec:ACE}

The ISDF decomposition can be combined with the recently developed
adaptively compressed exchange operator (ACE)
formulation\cite{JCTC_12_2242_2016_ACE, JCTC_13_1188_2017_ACE} to
further reduce the cost of hybrid functional KSDFT calculations.

In the ACE formulation, the operator $V_{X}$ is replaced by a
rank-$n$ operator $V_X^{\mathrm{ACE}}$ that satisfies
\[
V_{X}\psi_j = V_X^{\mathrm{ACE}}\psi_j
\]
for $j=1,2,...,n$, where $n$ is much less than the total number of grid
points $N_{g}$.  The operator
$V_X^{\mathrm{ACE}}$ can be written in the form
\begin{equation}
  V_X^{\mathrm{ACE}}(\vr,\vr') = -\sum_{k=1}^{n} \xi_{k}(\vr)\xi_{k}(\vr')
  \label{eqn:ACE}
\end{equation}
for some vectors $\{\xi_k\}$. The application of $V_X^{\mathrm{ACE}}$ to
a set of orbitals resembles the application of a nonlocal
pseudopotential operator.
This does not require any Poisson
solves, and is much cheaper than applying $V_X$ to these
orbitals in an iterative diagonalization procedure used to update
the set of occupied orbitals $\{\varphi_i\}$ in the SCF iteration.
However, the construction of $V_X^{\mathrm{ACE}}$, which must be
performed in each (outer) SCF iteration, still requires applying
$V_{X}$ to $\psi_j$ to produce
\begin{equation}
  W_{j}(\vr) = (V_{X}[\{\varphi_{i}\}]\psi_{j})(\vr) \quad
  j=1,\ldots,n.
  \label{eqn:ACEW}
\end{equation}
The basis vectors  $\{\xi_k\}$ that appear in Eq.~\eqref{eqn:ACE} are
obtained from $W_j$ via
\begin{equation}
 \xi_{k}(\vr) = \sum_{j=1}^{n} W_{j}(\vr)
  (L^{-T})_{jk},
 \label{eqn:Wscal}
\end{equation}
where $L$ is the lower triangular Cholesky factor of the matrix $M$.
The $(i,j)$th element of $M$ is given by
\begin{equation}
  M_{ij} = \int \psi_{i}(\vr) W_{j}(\vr) \ud \vr.
  \label{eqn:ACEM}
\end{equation}
Because $M$ is a symmetric negative definite matrix of size $n$,
a Cholesky factorization can be used to decompose $-M$ as $-M=LL^{T}$,
where $L$ is unit lower triangular. Since
Eq.~\eqref{eqn:ACEW} is performed only in each outer iteration,
which is less frequent than applying $V_{X}$ to $\psi_j$ in every
step of the diagonalization procedure, the use of ACE significantly
reduces the cost of hybrid functional calculation, without requiring
any approximation to the computation of the $W_{j}$'s.



The ISDF decomposition can be readily used to accelerate the
computation of $W_{j}$'s in Eq.~\eqref{eqn:ACEW}. However,
straightforward computation using in Eq.~\eqref{eqn:ACEM} may result
in an $M$ matrix that is not symmetric, let alone being negative
definite. To see this, we combine
Eq.~\eqref{eqn:applyVXDF} and Eq.~\eqref{eqn:ACEM}  to obtain
\[
  M_{ij} = \sum_{\mu=1}^{N_{\mu}}\left(\int \psi_{i}(\vr)
  P^{\varphi}(\vr,\vr_{\mu}) V^{\zeta}_{\mu}(\vr)\ud \vr\right)
  \psi_{j}(\vr_{\mu}),
\]
which may be different from
\[
  M_{ji} = \sum_{\mu=1}^{N_{\mu}}\left(\int \psi_{j}(\vr)
  P^{\varphi}(\vr,\vr_{\mu}) V^{\zeta}_{\mu}(\vr)\ud \vr\right)
  \psi_{i}(\vr_{\mu}).
\]
The lack of symmetry may result in numerical stability problems in
the subsequent Cholesky factorization of $M$. In order to overcome
this problem, we can apply the ISDF decomposition in a symmetric
fashion as follows. Note that
\begin{equation}
  M_{ij} = \sum_{l=1}^{N_{e}} \int \psi_{i}(\vr) \varphi_{l}(\vr)
  K(\vr,\vr') \varphi_{l}(\vr')\psi_{j}(\vr') \ud \vr \ud \vr'.
  \end{equation}
Hence, we can use ISDF to expand both the
$\varphi_{\ell}(\vr)\psi_i(\vr)$ and
$\varphi_{\ell}(\vr')\psi_j(\vr')$ pairs in terms of the
interpolating vector $\{\zeta_\mu\}$ to obtain
  \begin{equation}
  \begin{split}
  M_{ij} \approx & \sum_{l=1}^{N_{e}} \sum_{\mu,\nu=1}^{N_{\mu}}
  \left(\int \zeta_{\mu}(\vr) K(\vr,\vr') \zeta_{\nu}(\vr')\ud \vr \ud
  \vr'\right) \varphi_{l}(\vr_{\mu}) \varphi_{l}(\vr_{\nu})
  \psi_{i}(\vr_{\mu}) \psi_{j}(\vr_{\nu})\\
  =& \sum_{\mu,\nu=1}^{N_{\mu}}
  \left(\int \zeta_{\mu}(\vr) K(\vr,\vr') \zeta_{\nu}(\vr')\ud \vr \ud
  \vr'\right) P^{\varphi}(\vr_{\mu},\vr_{\nu})
  \psi_{i}(\vr_{\mu}) \psi_{j}(\vr_{\nu}) \\
  =& \sum_{\mu,\nu=1}^{N_{\mu}} \psi_{i}(\vr_{\mu})\wt{M}_{\mu\nu}
  \psi_{j}(\vr_{\nu}),
  \end{split}
  \label{eqn:ACEMSym}
\end{equation}
where
\begin{equation}
  \wt{M}_{\mu\nu} =
  \left(\int \zeta_{\mu}(\vr) K(\vr,\vr') \zeta_{\nu}(\vr')\ud \vr \ud
  \vr'\right) P^{\varphi}(\vr_{\mu},\vr_{\nu}).
  \label{eqn:Mtilde}
\end{equation}
Since both the first and second factors on the right hand side of
Eq.~\eqref{eqn:Mtilde} are symmetric, $ \wt{M}_{\mu\nu} $ is clearly
symmetric, which guarantees the symmetry and the definiteness of $M$ defined in
Eq.~\eqref{eqn:ACEMSym}.

In the symmetric formulation given by Eq.~\eqref{eqn:ACEMSym}, $M$
is automatically a symmetric negative definite matrix. Therefore,
the Cholesky factorization of $M$ yields the ACE operator according
to Eq.~\eqref{eqn:ACE}. Hence we refer to the combined method using
ACE and ISDF as the ACE-ISDF method.


\section{Parallel implementation}\label{sec:Parallel}

In this section, we demonstrate an efficient parallel implementation
of the ACE-ISDF method for hybrid functional calculations in a
planewave basis set.

%
%

\begin{figure}[htbp]
\begin{center}
\includegraphics[width=0.5\textwidth]{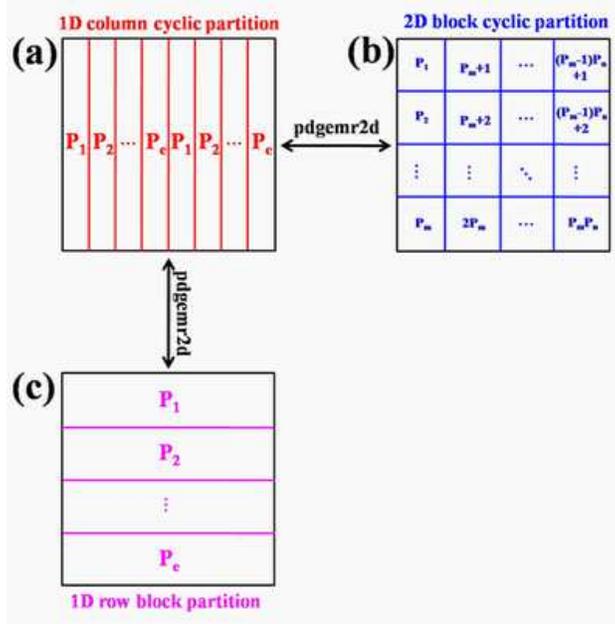}
\end{center}
\caption{(Color online) Three different types of data partition for
the matrix used in the ACE-ISDF formulation for hybrid density
functional calculations, (a) 1D column cyclic partition (1 $\times$
P$_\text{e}$ MPI processor grid), (b) 2D block cyclic partition
(P$_\text{m}$ $\times$ P$_\text{n}$ MPI processor grid) and (c) 1D
row block partition (P$_\text{e}$ $\times$ 1 MPI processor grid) .
P$_\text{e}$ is total computational cores used in the ACE-ISDF
formulation and P$_\text{m}$ $\times$ P$_\text{n}$ = P$_\text{e}$.}
\label{fig:ParallelPartition}
\end{figure}

Denote the matrix of the discretized orbitals by
$\Phi=[\varphi_1,\ldots,\varphi_{N_e}]$, and
$\Psi=[\psi_1,\ldots,\psi_{n}]$. When P$_{\text{e}}$ processors are
used, these orbitals are stored using the 1D column cyclic partition
as shown in Figure~\ref{fig:ParallelPartition}(a) so that the
application of the Hamiltonian operator (excluding the exchange
part) to the orbitals $\Psi$ can be easily parallelized.  In
particular, the Laplacian operator can be applied through the use of
a sequential Fast Fourier transformation (FFT) library. Moreover,
the application of the local and nonlocal pseudopotentials in the
real space representation is also rather straightforward. The
application of the ACE operator to $\Psi$ involves two matrix-matrix
multiplication operations, and can be done most efficiently by using
a row-based partition shown in Figure~\ref{fig:ParallelPartition}(c)
(see e.g. Ref.\cite{JCTC_13_1188_2017_ACE} for more details).
However, in the ACE-ISDF procedure,  a 2D block cyclic partition
shown in Figure~\ref{fig:ParallelPartition}(b) is the most efficient
data distribution scheme for performing a number of dense linear
algebra operations such as QRCP implemented in the ScaLAPACK
software package.\cite{ScaLAPACK_2011} The conversion among
different data storage formats is performed using the
\texttt{pdgemr2d} subroutine in the ScaLAPACK software
package.\cite{ScaLAPACK_2011}
\begin{figure}[htbp]
\begin{center}
\includegraphics[width=0.5\textwidth]{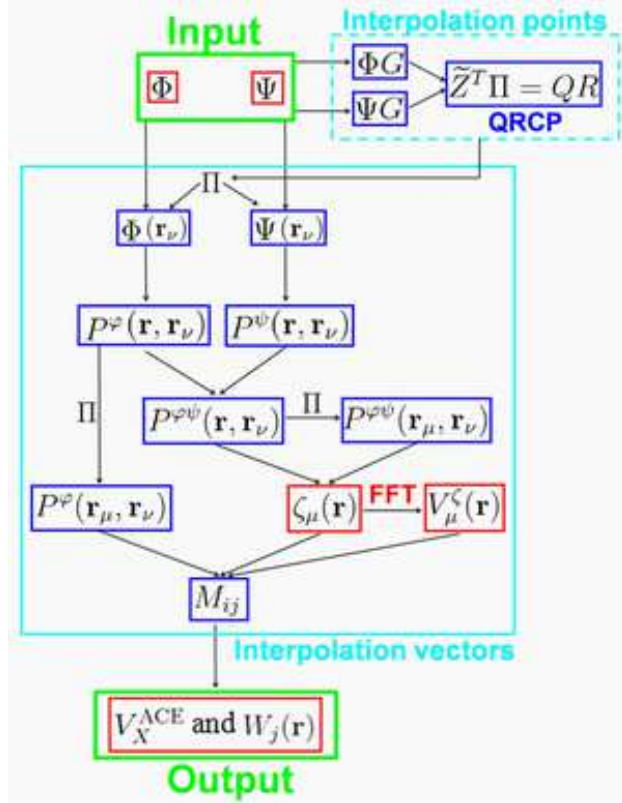}
\end{center}
\caption{(Color online) Flowchart of the ACE-ISDF formulation for
constructing the ACE exchange operator in PWDFT. Red and blue boxes
respectively represent 1D column cyclic partition and 2D block
cyclic partition for the matrices used in the ACE formulation. Once
the interpolation points $\{\vr_{\mu}\}_{\mu=1}^{N_{\mu}}$ are
obtained from QRCP at the first outer SCF iteration, the
interpolation vectors $\{\zeta_{\mu}(\vr)\}_{\mu=1}^{N_{\mu}}$ can
be updated directly from the input vectors at other outer SCF
iterations.} \label{fig:ParallelFormulation}
\end{figure}

More specifically, we describe the various quantities in the
ACE-ISDF method and the corresponding storage formats in
Figure~\ref{fig:ParallelFormulation}.  Starting from $\Phi$ and
$\Psi$ distributed in the column cyclic partition, we first
transform these matrices into the 2D block cyclic partition to
generate the $\wt{Z}$ matrix. We then perform the QRCP procedure to
obtain the permutation matrix $\Pi$. The interpolation points
$\{\hr{\mu}\}_{\mu=1}^{N_{\mu}}$ are retrieved from the permutation
matrix $\Pi$. This is a small vector of size $N_{\mu}$, and are
shared among all processors.

In order to construct the interpolation vectors, we distribute the
columns of the quasi density matrices $P^{\varphi}(\vr_k,\hat{\vr}_{\mu})$ and
$P^{\psi}(\vr_k,\hr{\mu})$
in a 2D block cyclic fashion so
that the matrix $ZC^T$ defined in \eqref{eq:newzct} can be evaluated
in parallel via local Hadamard multiplications.  The matrix $CC^T$
that appears in \eqref{eq:Theta} can be obtained by simply
subsampling rows of $ZC^T$.

The resulting discretized interpolation vectors
$\Omega=[\zeta_{1},\ldots,\zeta_{N_{\mu}}]$ can be obtained by
calling a ScaLAPACK linear equation solver. The 2D block cyclically
distributed solution is redistributed and partitioned by a 1D column
partition for computing the Coulomb-like potential for the
interpolation vectors
$V^{\zeta}=[V^{\zeta}_{1},\ldots,V^{\zeta}_{N_{\mu}}]$ as in
Eq.~\eqref{eqn:Vzeta}. $\{V^{\zeta}\}$ are then converted back
to the 2D block cyclic distribution pattern.  Finally, the $\{W_{j}\}$
in Eq.~\eqref{eqn:ACEW} can be computed using matrix-matrix
multiplication in the 2D partition as in Eq.~\eqref{eqn:applyVXDF},
and then converted to 1D  row partition.

In order to implement the symmetric formulation for the $M$ matrix
in Eq.~\eqref{eqn:ACEMSym} as required by ACE, we form the matrix
$\widetilde{M}$ in Eq.~\eqref{eqn:Mtilde} in parallel within a  2D
block cyclic distribution scheme, and the $M$ matrix in
Eq.~\eqref{eqn:ACEMSym} can be obtained by two parallel
matrix-matrix multiplication calls. Finally, we perform a parallel
Cholesky factorization of $M$ on the 2D block cyclic grid, and the
ACE vectors $\Xi=[\xi_{1},\ldots,\xi_{n}]$ are partitioned by rows
on a 1D processor grid.  This gives the $V_{X}^{\mathrm{ACE}}$
implicitly, and can be readily used in subsequent iterations.

\section{Numerical results} \label{sec:Result}



We demonstrate the performance of the ACE-ISDF method using the
DGDFT (Discontinuous Galerkin Density Functional Theory) software
package.\cite{JCP_231_2140_2012_DGDFT, JCP_143_124110_2015_DGDFT,
PCCP_17_31397_2015_DGDFT, JCP_145_154101_2016_DGDFT,
JCP_335_426_2017_DGDFT} DGDFT is a massively parallel electronic
structure software package designed for large scale DFT calculations
involving up to tens of thousands of atoms. It includes a
self-contained module called PWDFT for performing planewave based
electronic structure calculations (mostly for benchmarking and
validation purposes). We implemented the ACE-ISDF method in PWDFT.
We use the Message Passing Interface (MPI) to handle data
communication, and the Hartwigsen-Goedecker-Hutter (HGH)
norm-conserving pseudopotential\cite{PRB_58_3641_1998_HGH}. All
calculations use the HSE06
functional.\cite{JCP_124_219906_2006_HSE06} \REV{All calculations
are carried out on the Edison systems at the National
Energy Research Scientific Computing Center (NERSC). Each node consists
of two Intel ``Ivy Bridge'' processors with $24$ cores in total and
64 gigabyte (GB) of memory. Our implementation only uses MPI.
The number of cores is equal to the number of MPI ranks used in the simulation.}

\begin{figure}[htbp]
\begin{center}
\includegraphics[width=0.5\textwidth]{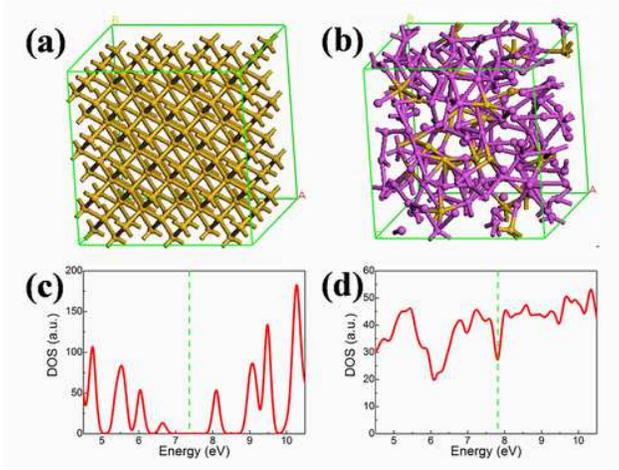}
\end{center}
\caption{(Color online) \REV{The atomics structures of (a) Si$_{216}$ and
(b) Al$_{176}$Si$_{24}$. The yellow and pink balls denote silicon
and aluminum atoms, respectively. The total densities of states (DOS) of
(c) semiconducting Si$_{216}$ and (d) metallic Al$_{176}$Si$_{24}$.
The Fermi levels are marked by green dotted lines.}} \label{fig:DOS}
\end{figure}

In this section, we demonstrate the accuracy of the ACE-ISDF method
for accelerating hybrid functional calculations, using a bulk
silicon system Si$_{216}$ and a disordered system
Al$_{176}$Si$_{24}$\cite{JCP_145_154101_2016_DGDFT} \REV{as shown in
Figure~\ref{fig:DOS} (a) and (b), respectively. The Si$_{216}$
system is semiconducting with an energy gap of $E_\text{gap} = 1.45$
eV, and the Al$_{176}$Si$_{24}$ system is metallic with
$E_\text{gap} < 0.1$ eV. The density of states of the two systems
are shown in Figure~\ref{fig:DOS} (c) and (d), respectively.} All
systems are closed shell systems, and the number of occupied bands
is $N_\text{band} = N_{e}/2$. We include two unoccupied bands for
computing the energy gap in the systems. We show the parallel
scalability of our implementation using 7 bulk silicon systems with
$64$ to $4096$ atoms.\cite{JCTC_13_1188_2017_ACE} Finally, we use
the ACE-ISDF method in a hybrid DFT calculation to study the
electronic structure of vacancy defect in a silicon supercell that
contains $1000$ Si atoms.

\subsection{Accuracy}\label{sec:Accuracy}


In the previous work,\cite{JCTC_12_2242_2016_ACE,
JCTC_13_1188_2017_ACE} we demonstrated that the ACE formulation can
significantly accelerate hybrid functional calculations without loss
of accuracy. Hence the results from the ACE calculation is used as
the baseline for comparison in assessing the accuracy of the
ACE-ISDF method. Table~\ref{Accuracy} shows the convergence of the
ACE-ISDF method as a function of the rank parameter $c$ for the
Si$_{216}$ and Al$_{176}$Si$_{24}$ systems. We measure the accuracy
in terms of the valence band maximum (VBM) energy level, the
conduction band minimum (CBM) energy levels, the energy gap, the
Hartree-Fock (HF) exchange energy, the total energy as well as the
atomic force. The energy cutoff $E_{\text{cut}}$ is set to $20$
Hartree. We define the errors in the HF energy, the total energy and
the atomic force respectively as
\[
\Delta{E_\text{HF}} = (E_\text{HF}^\text{ACE-ISDF} -
E_\text{HF}^\text{ACE})/N_{A},
\]
\[
\Delta{E} = (E^\text{ACE-ISDF} - E^\text{ACE})/N_{A},
\]
\[
{\Delta}F = \max_I|F_I^\text{ACE-ISDF} - F_I^\text{ACE}|.
\]
Here $N_{A}$ is the total number of atoms and $I$ is the atom
index.
%
\begin{table}
\caption{The accuracy of hybrid functional calculations (HSE06)
obtained by the ACE-ISDF method as a function of the rank parameter $c$
for Si$_{216}$ and Al$_{176}$Si$_{24}$. The unit for VBM ($E_\text{VBM}$),
CBM ($E_\text{CBM}$) and the energy gap $E_\text{gap}$ is eV.
The unit for the error in the Hartree-Fock exchange energy ${\Delta}E_\text{HF}$
and the total energy ${\Delta}E$ is Hartree/atom, and the unit for the error
in atomic forces ${\Delta}F$ is Hartree/Bohr. We use results from
the ACE-enabled hybrid functional calculations as the reference.}
\label{Accuracy}
\begin{tabular}{ccccccc} \\ \hline \hline
\multicolumn{7}{c}{ACE-ISDF: Semiconducting Si$_{216}$ ($N_\text{band} = 432$)} \ \\
\hline
$c$  &  $E_\text{VBM}$ & $E_\text{CBM}$ & $E_\text{gap}$ & ${\Delta}E_\text{HF}$ & ${\Delta}E$ & ${\Delta}F$   \ \\
\hline
4.0  & 6.5303   & 8.4367    & 1.9064    & 3.53E-03  & 4.13E-03  & 1.49E-03  \ \\
5.0  & 6.5923   & 8.3652    & 1.7729    & 2.37E-03  & 2.75E-03  & 1.30E-03  \ \\
6.0  & 6.6786   & 8.2535    & 1.5749    & 7.34E-04  & 1.06E-03  & 1.00E-03  \ \\
7.0  & 6.6893   & 8.1554    & 1.4661    & 3.56E-04  & 4.61E-04  & 6.61E-04  \ \\
8.0  & 6.6763   & 8.1341    & 1.4578    & 1.16E-04  & 1.64E-04  & 3.05E-04  \ \\
9.0  & 6.6652   & 8.1164    & 1.4512    & 4.74E-05  & 7.50E-05  & 1.48E-04  \ \\
10.0 & 6.6565   & 8.1085    & 1.4520    & 2.33E-05  & 4.11E-05  & 1.09E-04  \ \\
12.0 & 6.6487   & 8.1001    & 1.4514    & 7.27E-06  & 1.54E-05  & 5.88E-05  \ \\
16.0 & 6.6467   & 8.0959    & 1.4492    & 1.57E-06  & 2.92E-06  & 1.67E-05  \ \\
20.0 & 6.6466   & 8.0942    & 1.4476    & 5.40E-07  & 7.87E-07  & 5.54E-06  \ \\
ACE  & 6.6467   & 8.0934    & 1.4466    & 0.00E-00  & 0.00E-00  & 0.00E-00  \ \\
\hline \hline
\multicolumn{7}{c}{ACE-ISDF: Metallic Al$_{176}$Si$_{24}$ ($N_\text{band} = 312$)} \ \\
\hline
$c$  &  $E_\text{VBM}$ & $E_\text{CBM}$ & $E_\text{gap}$ & ${\Delta}E_\text{HF}$ & ${\Delta}E$ & ${\Delta}F$   \ \\
\hline
4.0  & 7.8907   & 7.9963    & 0.1056    & 7.20E-03  & 8.06E-03  & 8.96E-03  \ \\
5.0  & 7.8173   & 7.9103    & 0.0930    & 3.46E-03  & 3.76E-03  & 3.96E-03  \ \\
6.0  & 7.7810   & 7.8833    & 0.1023    & 1.33E-03  & 1.69E-03  & 2.32E-03  \ \\
7.0  & 7.7805   & 7.8742    & 0.0937    & 5.97E-04  & 6.41E-04  & 1.60E-03  \ \\
8.0  & 7.7717   & 7.8710    & 0.0993    & 1.90E-04  & 2.03E-04  & 5.55E-04  \ \\
9.0  & 7.7719   & 7.8710    & 0.0991    & 6.92E-05  & 7.44E-05  & 3.10E-04  \ \\
10.0 & 7.7713   & 7.8699    & 0.0986    & 3.20E-05  & 3.55E-05  & 1.53E-04  \ \\
12.0 & 7.7712   & 7.8695    & 0.0983    & 1.16E-05  & 1.38E-05  & 9.23E-05  \ \\
16.0 & 7.7704   & 7.8698    & 0.0994    & 3.26E-06  & 4.43E-06  & 4.34E-05  \ \\
20.0 & 7.7703   & 7.8695    & 0.0992    & 1.27E-06  & 1.93E-06  & 2.18E-05  \ \\
ACE  & 7.7701   & 7.8695    & 0.0994    & 0.00E-00  & 0.00E-00  & 0.00E-00  \ \\
\hline \hline
\end{tabular}
\end{table}

%

Our calculations show that the ACE-ISDF method can produce highly
accurate results with a moderate rank parameter $c$. (Recall that
the rank of the ISDF approximation is $N_{\mu}=c N_\text{band}$.)
The accuracy of the approximation can be improved systematically
by increasing the rank parameter $c$. \REV{When a relatively small $c$ value
(e.g., $c = 6.0$) is used, the error in the total energy of
both the Si$_{216}$ and the Al$_{176}$Si$_{24}$ systems is already
below that required to reach the chemical accuracy of 1 kcal/mol (1.6
Hartree/atom)\cite{RMP_71_1267_1999}.
} For Si$_{216}$, the errors in the HF energy, the total energy and
the atomic force systematically decrease from $\mathcal{O}(10^{-3})$
to $\mathcal{O}(10^{-6}\sim 10^{-7})$ when $c$ is adjusted from 6.0
to 20.0.
\REV{We note that the total energy convergence with respect to the rank
parameter is similar between Si$_{216}$ and Al$_{176}$Si$_{24}$.
The fact that the rank parameter $c$ is independent of
the band gap makes ISDF more attractive than linear
scaling methods\cite{JCP_105_2726_1996, JCP_109_1663_1998,
JCP_135_034110_2011, JPCA_114_1039_2010} whose accuracy is controlled
by the level of truncation in the density matrix, which in turn depends
strongly on the band gap.}

\subsection{Efficiency}\label{sec:efficiency}

We demonstrate the efficiency of the ACE-ISDF method by showing its
performance in a hybrid DFT calculation for a bulk silicon system
with $1000$ atoms ($N_\text{band}$ = 2000) on $2000$ cores.  In each
outer iteration, the cost of hybrid functional calculations consists
of the cost for constructing the ACE operator (with or without ISDF)
and the amount of work performed in the inner SCF iterations.

\REV{Table~\ref{Efficiency} shows the wall clock time spent in
major components of the ACE-ISDF and ACE calculations, respectively.
The main cost for constructing the ACE operator without using the
ISDF decomposition is in the solution of $\Or(N_{e}^2)$ Poisson-like
equations via FFTs. For this silicon system, the number of
Poisson-like equations to be solved in each outer iteration is as
large as $N_\text{bands}^2 = 4,000,000$. To show the detailed cost
of constructing the ACE operator using the ISDF method, we report
the timing measurements for selecting the interpolation points (IP),
computing the interpolation vectors (IV) and other linear algebra
operations and FFTs (labeled by `Other'). The IP selection is performed once
only in the first outer SCF iteration. Computing the IVs constitutes
a major part of the cost in the construction of the ACE operator in
subsequent outer SCF iterations. The time spent in solving Poisson
equations using FFTs, which we list in the parenthesis next to time
spent in the remaining parts of the ACE-ISDF calculation for comparison, is
negligibly small. The reason that this cost is so small is that the
use of ISDF significantly reduces the number of Poisson equations to
be solved from $N_\text{bands}^2 = 4,000,000$ to $N_{\mu} = 12,000$
(when the rank parameter $c$ is set to 6.)

\begin{table}
\caption{The wall clock time (in seconds) spent in the
components of the ACE-ISDF and ACE enabled hybrid DFT calculations
related to the exchange operator,
for Si$_{1000}$ on 2000 Edison cores at different $E_{\text{cut}}$
levels.  The corresponding number of real space grid points used to
represent the wavefunction is labeled by $N_g$.
We use the rank parameter $c$ = 6.0 in the ACE-ISDF
calculation.}
\label{Efficiency}
\begin{tabular}{ccccccccc} \\ \hline \hline
\multicolumn{2}{c}{Si$_{1000}$}      &  & & \multicolumn{3}{c}{ACE-ISDF}  & & ACE  \ \\
$E_{\text{cut}}$ &  $N_g$  & & &  IP  &  IV & Other (FFT)   & & FFT  \ \\
\hline
10  &  74$^3$  & &   & 50.22  & 11.13 & 8.56 (0.28) & & 101.10 \ \\
20  &  104$^3$ & &   & 105.95 & 24.52 & 20.52 (1.17) & & 148.73 \ \\
30  &  128$^3$ & &   & 222.36 & 40.67 & 32.88 (1.31) & & 302.98 \ \\
40  &  148$^3$ & &   & 454.42 & 63.56 & 54.95 (3.08) & & 807.31 \ \\
\hline \hline
\end{tabular}
\end{table}

When $E_{\text{cut}}$ is set to 10 Hartree and $c$ is set to 6.0,
the IP and IV computations take 50.22 s and 11.13 s respectively in
ACE-ISDF. The total amount of time spent in the construction of the
ACE operator via ISDF in the first SCF iteration, which is 70 s
(50.22+11.13+8.56), is already lower than that used to construct the
ACE operator without ISDF (a procedure dominated by solving a larger
number of Poisson-like equations via FFTs), which is roughly 101.10
s. In each subsequent outer SCF iteration, a total of 19.86 s
(11.13+8.56) are used to construct the ACE operator via ISDF.  This is
already comparable to the time of one SCF iteration in GGA
calculations, which is $17.89$ s.


Note that for some complex systems, more inner SCF iterations might be
required in each outer SCF iteration to reach convergence. For
example, for the disordered Al$_{176}$Si$_{24}$ system, we need to
use 14 inner SCF iterations per outer SCF iteration. As a result,
the cost difference between ACE and ACE-ISDF is magnified, and
ACE-ISDF is even more advantageous in these situation.

To illustrate the reduction of cost the ACE-ISDF scheme has achieved, we
report that the average time spent in the construction and application
of the exchange operator per outer SCF iteration in the conventional
hybrid functional calculations is 1146.36 s, which is nearly two orders
of magnitude higher than that used in ACE-ISDF. The large cost mainly
comes from the fact that conventional hybrid functional calculations
require solving $\mathcal{O}(N_e^2)$ Poisson like equations using
FFTs in each step of an iterative diagonalization procedure (e.g. PPCG)
when the exchange operator is applied to a set of $\mathcal{O}(N_e)$ orbitals.
For this system, on average $17$ such operations need to be performed
during each outer iteration.
The ACE formulation reduces the cost by only requiring the
exchange operator to be applied once per outer iteration, and the
ACE-ISDF method further reduces the cost for this single application of
the exchange operator.


As we discussed earlier, since IP calculation is significantly more
expensive, it is important to perform such a calculation only
once in order to make ACE-ISDF efficient. We found that
the Hartree-Fock exchange energy $E_\text{HF}$ obtained from using a
fixed set of IPs throughout the SCF iteration differs only slightly
from that obtained by recalculating IPs in each outer SCF iteration
(dynamic IP) as shown in Table~\ref{tab:IP} for Si$_{216}$ and Al$_{176}$Si$_{24}$.
We can clearly see from the table that the difference between the first and
the second columns is much smaller than the difference between
the second and third columns. This shows that using fixed IP
is well justified.
\begin{table}
\caption{A comparison among Hartree-Fock exchange energies computed
by ACE only and by ACE-ISDF with a fixed or changing set of IPs. The
rank parameter $c$ used in ACE-ISDF is set to 6.0 for the Si$_{216}$
and Al$_{176}$Si$_{24}$ systems.} \label{tab:IP}
\begin{tabular}{cccccc} \\ \hline \hline
Systems     & &    fixed IP in ACE-ISDF   &  dynamic IP in ACE-ISDF  & & ACE  \  \\
\hline
Si$_{216}$  & &     -45.3225              &  -45.3250 & & -45.4835 \ \\
Al$_{176}$Si$_{24}$  & &     -25.9112              &  -25.9093 & & -26.1745 \ \\
\hline \hline
\end{tabular}
\end{table}
Hence in practice we only use the QRCP decomposition in the first outer
SCF iteration to select IPs, whereas the update of the
basis vectors $\{\zeta_{\mu}(\vr)\}_{\mu=1}^{N_{\mu}}$ is performed in
each subsequent outer SCF iterations.}





Table~\ref{Efficiency} also shows that, as $E_{\text{cut}}$ is increased from
10 to 40 Hartree, the time spent in constructing the ACE operator increases
from 101.10 s to 807.31 s. The time spent in identifying the
interpolation points via QRCP also increases from 50.22 s to 454.42
s. However, the time used to compute the interpolation vectors only
increases from 11.13 s to 63.56 s. Given that the interpolating
points only need to be selected once, the performance gain
achieved by ISDF is more notable at a larger $E_{\text{cut}}$ value.
This feature is particularly attractive for calculations that
requiring large kinetic energy cutoff, such as those involving
transition metal oxides.



\subsection{Parallel scalability}\label{sec:scalability}

\begin{figure}[htbp]
\begin{center}
\includegraphics[width=0.5\textwidth]{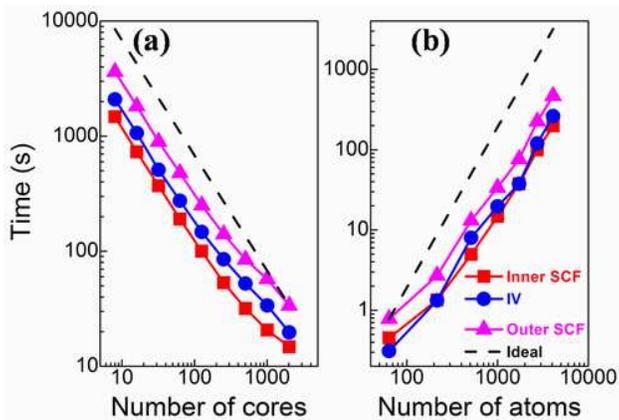}
\end{center}
\caption{(Color online) (a) The change of wallclock time in one
outer SCF iteration with respect to the number of cores for the
Si$_{1000}$ system (strong scaling). (b) The change of wallclock
time with respect to system size (weak scaling) on 8,192 cores.
The black dotted lines represent the ideal scaling.}
\label{fig:ParallelEfficiency}
\end{figure}

To illustrate the strong parallel scalability of the ACE-ISDF method
for large-scale hybrid DFT calculations, we report the change of the
wallclock time in one outer SCF iteration with respect to the number
of cores for the Si$_{1000}$ system. The measured wallclock time
includes time spent in one inner SCF iteration
and in the ACE-ISDF operator construction. We also report the weak
scaling of ACE-ISDF by showing the variation of the wallclock time
with respect to the system size for a calculation that uses 8,192
cores. These results are given in
Figure~\ref{fig:ParallelEfficiency} which shows that our implementation
of the ACE-ISDF method scales nearly perfectly up to 2,000 cores for
the Si$_{1000}$ system.
It also shows ACE-ISDF scales well
with respect to the system size (up to 4096 atoms) on 8192 cores.

\subsection{Application to vacancy defect in silicon}\label{sec:application}
%
%

Silicon is one of the most important materials in industry due to
its remarkable properties and a wide range of applications in
electronics. However, the presence of defects can significantly
affect these properties. Furthermore, defects also can be extremely
useful for designing innovative electronic
devices.\cite{RPP_45_1163_1982, MSSP_3_227_2000} Therefore, accurate
description of the electronic structures of silicon
defects\cite{RMP_86_253_2014} is required to examine their effects
on the electronic devices.

It is well known that DFT calculations based on LDA and GGA
functionals are not reliable in predicting electronic structures of
nanosystems.\cite{JCP_123_174101_2005} In particular, such DFT
calculations tend to underestimate the energy gap of semiconductors.
For example, GGA calculations that use the PBE
functional\cite{PRL_77_3865_1996_PBE} give an energy gap of 0.69 eV
in silicon, which is much smaller than that in bulk silicon measured
in the experiments (1.17 eV)\cite{APL_80_4834_2002}. The use of
hybrid functionals can mitigate this type of error.

The defect concentration in silicons used in experimental studies is
about 10$^{18}$ cm$^{-3}$. To faithfully represent experimental
conditions and avoid the nonphysical interactions between a defect
and its images introduced by periodic boundary conditions, a large
unit cell containing thousands of silicon atoms is required to in a
computational study. Systems of this size is beyond the capability
of existing planewave DFT software when a hybrid functional is used
in the calculation.  In this section we employ the ACE-ISDF method
implemented in PWDFT to calculate the energy levels of a vacancy
defect in an 1000-atom silicon system using the HSE06 hybrid
functional. The defect concentration is about 5 $\times$ 10$^{19}$
cm$^{-3}$ in this case, which is close to concentration used in
experimental studies. Our calculation based on the HSE06 functional
yields an energy gap of 1.28 eV, which is very close to the
experimental value of 1.17eV. We also find that a calculation that
uses a smaller unit cell that contains 512 atoms yields an energy
gap of 1.32 eV\cite{JCTC_13_1188_2017_ACE}, and hence the size
effect plays an important role.

\begin{figure}[htbp]
\begin{center}
\includegraphics[width=0.3\textwidth]{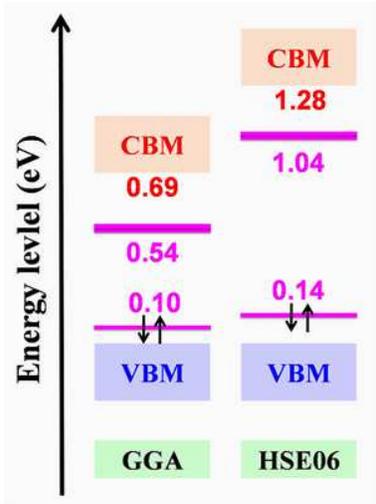}
\end{center}
\caption{(Color online) A comparison of VBM, CBM and defect energy
levels of a 1000-atom silicon system that contains a vacancy defect.
Two different types of exchange-correlation functionals (GGA-PBE and
HSE06) are used for geometry optimization and for computing the
electronic structure. All the energy levels are referenced to the
VBM energy, which is set to zero. The black arrows are used to mark
the occupied energy levels.} \label{fig:Si1000EnergyGap}
\end{figure}

Figure~\ref{fig:Si1000EnergyGap} shows the electronic structure of
the vacancy defect in the Si$_{1000}$ system computed with two
different exchange-correlation functionals (GGA-PBE and HSE06).
\REV{We
fully relax the structures respectively with the GGA and HSE06
exchange-correlation functionals by using the steepest descent algorithm
with the Barzilai-Borwein line search method~\cite{BarzilaiBorwein1988}
to optimize the geometry of the atomic configuration.
The optimal unit cell lattice constants we obtained are 5.46 and 5.45
{\AA} respectively for the GGA-PBE and HSE06
exchange-correlation functional based calculations. These
values are close to those reported in previous theoretical
studies.\cite{JPCM_24_145504_2012} }

Our DFT calculations show that hybrid
functional HSE06 calculations accurately describe the VBM and CBM
energy levels of the silicon and the defect energy levels introduced
by the vacancy defect. Furthermore, we can clearly see that the
vacancy defect introduces three defect states into the intrinsic
energy gap (1.28 eV) of silicon. These include a single defect state
$a_1$ and a doubly degenerate state $e$, 0.14 and 1.04 eV above the
VBM energy of silicon, respectively. The GGA based DFT calculations
fail to accurately predict these defect states.
%


\section{Conclusion} \label{sec:conclusion}


In this paper, we demonstrate that the interpolative separable
density fitting (ISDF) decomposition can be used to reduce the cost
hybrid functional calculations for large systems in a planewave DFT
code. The reduction in cost results from the construction of a set
of $cN_{e}$ numerical auxiliary basis vectors, where $c$ is a modest
constant. Using these auxiliary basis vectors instead of $N_{e}^2$
products of the occupied orbitals, we only need to solve
$\Or(N_{e})$ Poisson-like equations instead of $\Or(N_{e}^2)$
equations to apply an approximate exchange operator to a set of
occupied orbitals. The accuracy of the approximation depends
entirely on the rank parameter $c$, and we find that the choice of
$c$ is insensitive with respect to the band gap.

The ISDF decomposition can be performed in $\Or(N_{e}^3)$
operations. The interpolation points are chosen by a randomized QR
factorization with column pivoting (QRCP). It is relatively
expensive compared to other parts of the ISDF calculation. However,
this procedure only needs to be carried out once for all during the
first outer SCF iteration in a hybrid functional calculation. The
interpolation vectors can be computed via a least squares fitting
procedure that makes use of the separable nature of the functions to
be fitted.  The complexity of this step still scales as
$\Or(N_{e}^{3})$ but with a significantly smaller preconstant
compared to the cost of applying the uncompressed exchange operator
or the cost of QRCP. We are currently also exploring other methods
for selecting interpolation points that avoid the use of the QRCP
procedure, especially in the context of geometry optimization and
\textit{ab initio} molecular dynamics simulation.

For a moderate choice of rank parameter, the error in the total
energy per atom and the force can be kept under $10^{-3}$
Hartree/atom and $10^{-3}$ Hartree/Bohr respectively, for both
semiconducting and metallic systems. Meanwhile the computational
time can be reduced by up to an order of magnitude for applying the
exchange operator once to all Kohn-Sham orbitals. We demonstrated
that the ISDF decomposition can be combined with the adaptively
compressed exchange operator (ACE) formulation to reduce the cost of
ACE operator construction. The resulting ACE-ISDF method exhibits
excellent parallel scalability on high performance computers, and
significantly reduces the time required to perform hybrid functional
calculations by nearly two orders of magnitude. In particular, the
time spent in ACE-ISDF enabled hybrid functional calculation is only
marginally higher than that spent in DFT calculations that use local
and semilocal functionals.

However, we also find that hybrid functionals calculations often
require more iterations to converge compared to GGA calculations.
One main reason is the two level SCF iteration structure in hybrid
functional calculations, which may be inefficient especially in the
context of \textit{ab initio} molecular dynamics simulation. Further
reduction of the number of SCF iterations may close the final gap
between hybrid functional calculations and calculations with local
and semilocal functionals, and thus opens the door to the accurate
simulation of a vast range of nanomaterials using hybrid functionals
beyond reach today.

\section{Acknowledgments}

This work was partly supported by the Scientific Discovery through
Advanced Computing (SciDAC) program funded by U.S.~Department of
Energy, Office of Science, Advanced Scientific Computing Research
and Basic Energy Sciences (W. H., L. L. and C. Y.), by the National
Science Foundation under grant DMS-1652330 (L. L.), and by the
Center for Applied Mathematics for Energy Research Applications
(CAMERA) (L. L. and C. Y.). The authors thank the National Energy
Research Scientific Computing (NERSC) center and the Berkeley
Research Computing program at the University of California, Berkeley
for making computational resources available to them. We would like
to thank Francois Gygi, Jianfeng Lu, Meiyue Shao and Lexing Ying for
helpful discussions.

\footnotesize{

}

\vspace{3ex}

\[
\includegraphics[width=0.5\textwidth]{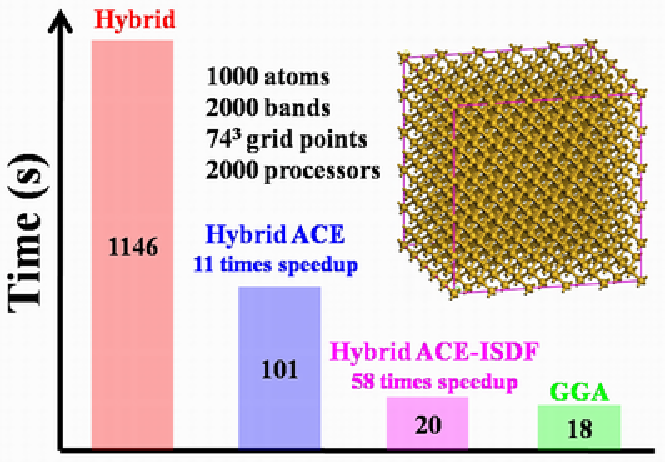}
\]
\centerline{TOC graphic}

\end{document}